# The *COVID19Impact* Survey: Assessing the Pulse of the COVID-19 Pandemic in Spain via 24 Questions

Nuria Oliver[1,2,3], Eng, PhD; Xavier Barber[4], MSc, PhD; Kirsten Roomp[5], BSc, PhD; Kristof Roomp[6]

[1]The European Laboratory for Learning and Intelligent Systems (ELLIS), Alicante, Spain

[2]The Spanish Royal Academy of Engineering, Madrid, Spain

[3]Commissioner for the President of the Valencian Region of Spain on Artificial Intelligence and COVID-19

[4]Center for Operations Research. Miguel Hernandez University, Alicante, Spain

[5]University of Luxembourg, Luxembourg

[6]Microsoft, Redmond, WA, US

**Corresponding Author:**
Nuria Oliver, Eng, PhD
The European Laboratory for Learning and Intelligent Systems, Alicante unit
University of Alicante
Alicante, Spain
Phone: 34 630726085
Email: nuria@alum.mit.edu

## *Abstract*

**Background**: Spain has been one of the most impacted countries by the COVID-19 pandemic. Since the first confirmed case of COVID-19 reported on January 31st, 2020, over 240,000 cases have been reported in Spain, resulting in over 27,000 deaths. The economic and social impact of the COVID-19 pandemic is without precedent. In this context, it is of paramount importance to quickly assess the situation and perception of citizens during the pandemic. Large-scale, online surveys have been shown to be an effective tool to carry out such rapid assessments.

**Objective:** The objective of the research described in this paper is to quickly assess the Spanish citizens' situation and perception on four areas related to the COVID-19 pandemic: their social contact behavior during the confinement, their personal economic impact, their workplace situation and their health status. We report overall statistics, carry out bi-variate statistical analysis and identify significant age and gender differences in people's situations and behavior during the COVID-19 pandemic in Spain. Moreover, we perform multi-variate logistic regression models and build a linear regression COVID-19 prevalence model using two of the questions of the survey and validate it with data from a seroprevalence study in Spain.

**Methods:** We obtained a large sample using an online survey with 24 questions related to COVID-19 for rapid and effective distribution. The self-selection online survey method of nonprobability sampling was used to recruit 156,614 participants via social media posts that targeted the general adult population (aged > 18 years old).

**Results**: Regarding the social behavior during confinement, we found that participants mainly left their homes to satisfy basic needs, such as go to the pharmacy, supermarket, and bakery (47.8%) and to go to work (31.3%). On average, 10.1% of respondents did not leave their homes and stayed at home during their confinement. We identified several statistically significant ($P<0.001$) differences in the social behavior across genders and age groups: 14.8% of female respondents vs 6.5% of male respondents reported not leaving their home; 26% of female participants vs 36.7% of male participants reported leaving to go to work and respondents aged 60 and older were almost twice as likely to stay home than younger participants (14.9% vs 7.6%). Individual transportation was largely preferred over other means of transportation (84.5%). The citizens' solidarity with the measures and resilience regarding the confinement is evident: most respondents (46.6%) believed that the government should implement more measures and 44.1% of participants reported being able to remain in confinement for one additional month. The survey answers reveal a significant economic impact of the pandemic in small businesses: 47.3% of respondents working in small (1-9 workers) companies reported having been financially affected and 19.4% reported facing bankruptcy at their work. We also find that economic impact is a key driver for resilience towards the confinement measures: a multi-variate logistic regression model reveals that those who report not having enough money to buy food have more than twice the probability of reporting that they could not stay in confinement for more than 1 week (OR=2.23, 95%CI[1.81, 2.77]). In terms of the ability to implement an effective quarantine, 27.2% of participants reported not having the necessary resources to isolate themselves. Regarding symptom prevalence, 16.8% of respondents reported having at least one COVID-19-related symptom and 7.1% reported having at least one of the most severe symptoms (fever, dry cough and difficulty breathing). The answers to the survey also point out to a lack of tests, with 6.1% of participants reporting that their doctors recommended they get tested, but no tests were available. From the symptoms and social contact behavior, we built a regression model to infer COVID-19 prevalence in Spain and report prevalence figures that are near or within the margin of error of a recent seroprevalence study carried out by the Spanish Ministry of Health. Moreover, a large portion of respondents who had tested positive (80.9%) reported having had close contact with an infected individual (e.g. friend or relative, patient, colleague, or client). Given such a large sample, the C.I. for a 95% confidence level is of ±0.843 for all reported proportions.

**Conclusions**: The Spanish population has shown high levels of compliance with confinement measures and resilience during the confinement period. In fact, at the time when the survey was deployed, most of the population demanded more measures. Close contacts play an important role in the transmission of the disease, particularly during confinement. The economic impact of the pandemic is evident in small companies. Gender and age matter regarding the social contact behavior, the economic and labor impact, and the ability to self-isolate. Quarantine infrastructure might be needed as over one quarter of the population reports lacking the necessary means to isolate themselves. The number of COVID-19 infected individuals is larger than the officially reported figures and can be estimated from the answers to our survey. During the early period of the pandemic, our survey also reveals a lack of tests and a significant difference ($P<0.001$) in attitudes towards testing availability between those with symptoms vs those without.

## Introduction

**Background**

The first cases of the coronavirus disease (COVID-19) were reported in Wuhan, China in December 2019. Since then, it has spread to 213 countries and territories, infecting over 6.2 million people and causing over 372,000 deaths worldwide as of June 1st, 2020 [1]. COVID-19 has caused significantly more infections and deaths, compared with previous outbreaks of Severe Acute Respiratory Syndrome (SARS) and Middle East Respiratory Syndrome (MERS), with an average infection rate of 2-2.5 people. The World Health Organization (WHO) declared a global COVID-19 pandemic on March 11, 2020 and to date has been unable to predict the duration of the pandemic [2].

The first confirmed case of COVID-19 in Spain was reported on January 31st, 2020 when a German tourist tested positive in the Spanish Canary Islands. However, this was an isolated, imported case. It was not until February 24th when Spain confirmed several new COVID-19 cases related to a recent COVID-19 outbreak in the North of Italy. Since that date, the number of COVID-19 cases grew exponentially in Spain, so that by March 30th, 2020 there were over 85,199 confirmed cases, 16,780 recoveries and the staggering figure of 7,424 deaths, according to the official figures. On March 25th, 2020, the death toll attributed to COVID-19 in Spain surpassed that of mainland China and it was only surpassed by the death toll in Italy. The economic and social impact of the COVID-19 pandemic in Spain is without precedent.

To combat the pandemic, the Spanish Government implemented a series of social distancing and mobility restriction measures. First, all classes at all educational levels were cancelled in the main hotspots of the disease: on March 10th, in the Basque Country, and on March 11st in the Madrid and La Rioja regions. All direct flights from Italy to Spain were cancelled on March 10th. On March 12th, the Catalan Government quarantined four municipalities that were particularly affected by the virus. On March 13th, the Government of Spain declared a state of emergency for two weeks across the entire country, which was later extended until April 11th and then renewed on a biweekly basis until June 21st. Unfortunately, different regions implemented containment measures at different times while still allowing travel to other regions, which might have enabled infected individuals to spread the virus. Since the state of emergency was established, all schools and university classes were cancelled; large-scale events and non-essential travel were forbidden, and workers were encouraged to tele-work. Despite these efforts, the daily growth rate in the number of confirmed COVID-19 cases continued to grow. Thus, on March 30th new mobility restriction and social distancing measures were implemented: all non-essential labor activity was to be interrupted for a 2-week period.

These interventions put a halt to the daily lives of most of the people in Spain. However, the number of confirmed cases, intensive care patients and deaths continued to grow exponentially. It is unclear how effective these measures will eventually be, as well as their impact on people's economic, physical, and mental well-being.

Given the speed of growth of the confirmed COVID-19 cases, rapid assessments of the population's situation and perceptions of the infection are of paramount importance. Traditional methods, such as population-representative household surveys are slow to design and deploy [3]. Phone surveys are generally faster to conduct, yet they are very labor-intensive, and often yield very low response rates (as low as 10% of less [4]). Moreover, the resulting sample might be very biased and difficult to reweight [5]. Given the limitations of these traditional methods and given the need for rapid data collection, large-scale online surveys are a valuable method to quickly assess and longitudinally monitor the situation and perceptions of the population in the context of a pandemic. Thus, to shed light on important, yet unknown questions related to COVID-19, we designed a 24-question online survey, called the *Covid19Impact* survey, to be deployed to the Spanish population. The survey was extremely well-received in Spain, becoming viral in 12 hours after its publication and yielding over 140,000 answers. It is one of the largest surveys in the world carried out in the context of the COVID-19 pandemic.

**Citizen Surveys during the COVID-19 Pandemic**

Other efforts to collect data from citizens regarding the COVID-19 pandemic have been deployed in multiple countries. The largest study to date involved the *Methods* smart-phone app, where 2,618,862 participants who self-reported symptoms on in the US and UK [6]. The study asked 40 questions focused on risk factors and symptoms, and published a predictive model based on risk factors and symptoms. In Canada, *FLATTEN* [7] has gathered data from 442,458 respondents as of June 1st 2020 and asks nine simple health and demographic related questions in order help monitor the spread of the virus in an anonymous manner. It allows the public to track COVID-19 cases on an interactive map in real-time. This is followed by the *International Survey on Coronavirus*, run by researchers at Harvard, Cambridge, IESE, and Warwick University, which has collected approximately 113,000 responses worldwide as of June 1st, 2020 [8]. It asks 18 questions concentrating on the psychological impact of the crisis. There were three main findings from the analysis of the answers of this survey: many respondents found their citizens and governments response to the COVID-19 pandemic was insufficient, this insufficient response was associated with lower mental well-being, and a strong government response was associated with an improvement in respondents view of their fellow citizens, government and a better mental well-being. The *COVID-19:CH Survey* in Switzerland, which aims to collect personal data related to COVID-19 testing with additional health and potential exposure related information. As of June 1st, 2020, ~12,800 surveys have been filled out [9]. The data collected is presented to the public in a visual format, giving information on, among other things, demographics, co-morbidities and symptoms. In Israel, the Weizmann Institute and the Ministry of Health are collecting data on basic demographics, health and potential exposure. Respondents were asked to fill the survey out on a daily basis for each family member [10]. The project tries to predict the location of COVID-19 outbreaks by analyzing information collected about the virus symptoms and public behavior in real-time. As of March 23rd, 2020 there were ~74,000 responses, a more current participant number is not available [11]. In Iran, a recent survey gathered data from 10,069 participants, asking questions related to demographics and health providing interesting findings regarding olfactory dysfunction in relation to COVID-19 [12].

Numerous efforts with smaller numbers of respondents have also taken place or are ongoing. In China, an early study was conducted between January 27th and February 1st, 2020 which relied on the Chinese social media and traditional media outlets, asking about knowledge, attitudes, and practices towards COVID-19 with 12 questions and receiving 6910 completed surveys [13]. There were numerous findings, including that most respondents felt that China could with the battle against the virus. An early international project was run from February 23rd to March 2nd, 2020 and collected data from the UK and the US utilizing an online platform managed by Prolific Academic Ltd and asked for knowledge and perceptions of COVID-19 using a convenience sample of 3000 participants to respond to 22 questions [14, 15] . The survey thus provided potential information to guide public health campaigns. In mid-March over 48 hours (March 14-16, 2020), 9009 responses were collected in the US; the 21 question survey had been posted on 3 social media platforms (Twitter, Facebook and Nextdoor), and collected data on symptoms, concerns and individual actions [16]. They showed that 95.7% of respondents made lifestyle changes, including handwashing, avoiding social gatherings, social distancing, etc. In the UK, data from 2,108 individuals was collected from March 17-18, 2020 attempting to identify sociodemographic adoption of social-distancing measures, ability to work from home, and both the willingness and ability to self-isolate [17], providing potential information to policy makers. From 26th to 29th March, 2020 an online survey (*FEEL-COVID*) used the snowballing method to collect data from 1106 respondents in India (453 responses being excluded due to being incomplete). The survey applied the Impact of Event-revised (IES-R) scale which measures psychological impact and found that almost one third of respondents were negatively psychologically impacted by the pandemic [18]. In the United States, the *COVID-19 Risk Survey* launched by Englander Institute for Precision Medicine, which had 3860 respondents on May 31st, 2020 asking demographic and health related questions [19]. The data collected is shared with the public daily in a visual, interactive format. Finally, the COVID-19 Screening Tool [20] hosted by Apple, in partnership with The Centers for Disease Control and Prevention (CDC) asks for demographics, health and travel related questions but does not publicize its results.

Our work complements these previous, related efforts, by focusing on Spain (one of the most affected countries by the COVID-19 pandemic) and by addressing four areas of people's experiences during the confinement: their social contact behavior, economic impact, labor situation and health status.

**This Study**

Despite the availability of data regarding the number of confirmed COVID-19 cases, hospitalized and intensive care patients and deaths, there is a scarcity of high-quality data about important questions related to the population's experience of the COVID-19 pandemic.

First, there is the issue of the under-reporting of confirmed cases and COVID-19 related deaths. Work by the Imperial College COVID-19 Response Team [21], estimated that 15% of the Spanish population could be infected by COVID-19. However, this figure was estimated to be much lower at around 5.3% by the preliminary results of a seroprevalence study carried out by the Spanish Ministry of Health [22]. Assessing the percentage of infected individuals is of utmost importance to build accurate epidemiological models and to assist policymakers in their decisions.

Second, there are unknowns regarding the sources of infection. Are people being infected by friends, family members, relatives, and co-workers? Or are they being infected because of serendipitous interactions in supermarkets or at the bakery? The effectiveness of different government interventions will depend on the answers to these questions.

Third, the economic impact that the COVID-19 crisis will have on people's lives is yet to be quantified. According to the latest figures from the Spanish Industry, Commerce and Tourism Ministry (January 2020), only 0.2% of Spanish companies have 250 or more employees; 44.6% of companies are micro (1-9 employees) or small (10-49 employees) and 54.4% of companies consist of the self-employed [23]. Small businesses are generally unprepared to confront such a crisis. Moreover, tourism represents 14.6% of Spanish GDP and 2.8 million of jobs and these are threatened by the COVID-19 pandemic [23]. Measuring the impact that COVID-19 is having on people's finances is of great value to policymakers. Finally, there is the personal experience related to having to be confined in the home for weeks. How much longer are citizens able to sustain this situation?

In this paper, we describe the *Covid19Impact* survey which was designed to answer the questions above. We present the methodology that we followed to gather a representative sample via a large-scale online survey, followed by the results of the analysis of the answers and the main insights derived from them. Finally, we describe our conclusions and lines of future work.

## *Methods*

**Sampling and Data Collection**

To answer the previously formulated questions, we designed a 24-question anonymous online survey that we refer to as the *Covid19Impact* survey, shown in (Multimedia Appendix 1). The survey is divided in 4 sections that address the different dimensions related to the citizens' experience during the COVID-19 crisis: their social contact in the last two weeks, the economic impact of the pandemic, their workplace situation and their health status. Moreover, the survey collects basic demographic (age range, gender, postal code) and home data (type of home and number/ages of people in the home)

We used the self-selection online survey method of non-probability sampling to recruit participants via social network posts (mainly Twitter and WhatsApp), asking the Spanish population (aged older than 18) to answer the survey. This sampling method is particularly suitable during a confinement situation where the mobility and social contact of the population is greatly reduced. Thus, the online distribution of the survey enabled fast access to it by large numbers of people.

In addition to distributing the survey on Twitter and WhatsApp, we used snowball sampling [24]. The goal was to collect as representative of a sample as possible in a short amount of time, as the COVID-19 situation is rapidly evolving, and new government measures might be implemented. The objective was to gather a snapshot of people's experiences regarding the four sections described above.

Anticipating the start of new mobility restriction and social distancing measures on Monday, March 30th, we deployed the survey on Saturday, March 28th at 8 PM. Via social media (Twitter and WhatsApp) and snowball sampling, we distributed the survey to a wide set of highly connected users who, in turn, distributed it to their contacts. The survey was also distributed by professional organizations, town halls, civil groups and associations. It inspired tens of thousands of citizens to not only contribute their own answers, but to share it with their friends, relatives, colleagues, and followers. In the 12 hours that followed, the survey went viral in Spain and by the afternoon of Monday, March 30th, we had collected over 140,000 answers. Figure 1 illustrates the growth in the number of answers over time and the peak was reached in the time frame between 4 PM – 5 PM on Saturday, March 29th, with more than 15,000 answers in one hour.

The initial version survey was delivered via Google Forms, which allowed us to write and deploy the survey in an anonymous, scalable and free manner within hours. The URL to the Google Forms was shared via bit.ly, such that we could estimate how many times the link had been shared. After reaching 140,000 answers, we began to hit scale limitations in Google Forms, so on March 30th, 2020 we moved the survey to Survey123 [25] for future editions of the data collection.

**Questionnaire Structure**

All questions are anonymized to preserve privacy and no personal information is collected. In addition, the snowball sampling methodology ensures anonymity and the absence of constraining or biasing factors as everyone contributed in a voluntary, and in many cases very committed, way. The survey can be found online in [26].

First, the survey obtains explicit consent from the users. Only when consent is granted and respondents confirm they are adults, respondents can respond to the rest of the questions.

The first section (Q1-Q4) gathers basic demographics: country, age range, gender and postal code. Next, there are 3 questions (Q5-Q7) related to the home situation: type of home, number of people in the home and their ages. The following 7 questions (Q8-Q14) address the social contact behavior of the respondents during the last two weeks. This is an important section of the survey as we aim to understand the level of social interaction that people have despite the social distancing measures. The questions ask about having had contact with infected individuals, whether children are taken care of outside the home, if they have an external person coming to their house (e.g. house cleaner), for what types of activities have they left their home and what transportation means have they used. The last two questions intend to capture people's perceptions of the confinement measures: if they think they are enough to contain the pandemic and for how long they would be able to tolerate the containment situation. Personal economic impact is assessed with questions Q15 and Q16, followed by three questions (Q17-Q19) related to their workplace situation. Finally, the last 5 questions (Q20-Q24) address their health state to assess how many people might be infected by the virus, determine the ability of citizens' to self-isolate and collect feedback regarding testing availability and testing results.

None of the questions, except for the consent question, are compulsory and all the health-related questions include "I prefer not to answer" as a choice.

**Figure 1.** Evolution of the number of answers collected by the *Covid19Impact* survey in the first two days since its launch, reported in 1-hour intervals

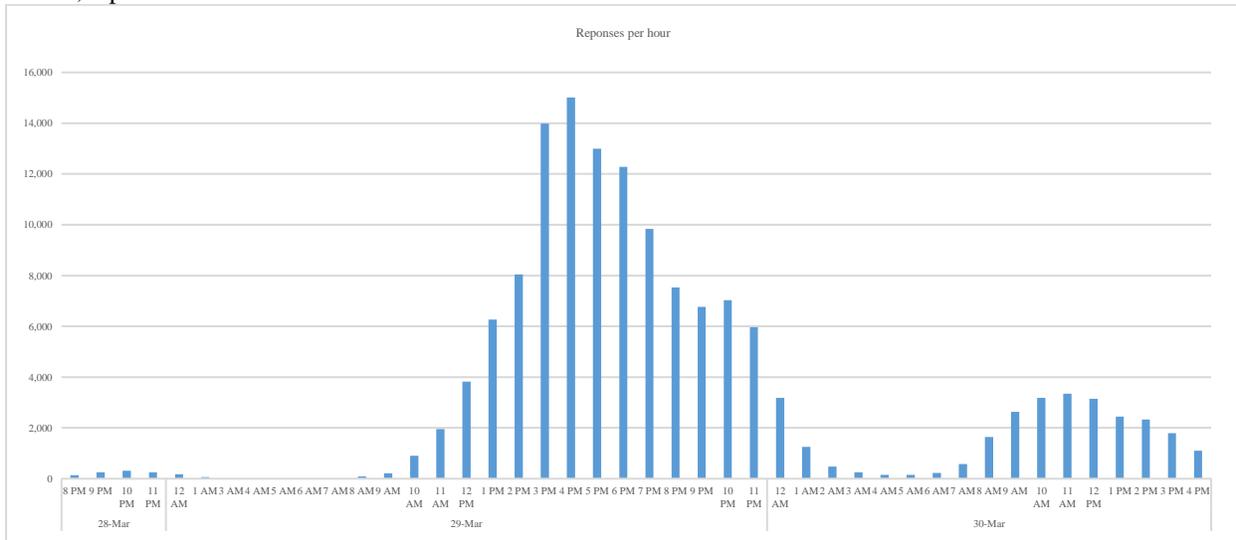

**Credibility and Validity**

Before widely deploying the survey, we carried out a pilot study to validate its content and proper anonymization with a small sample of participants. The questions were written in Spanish and English. Once all bugs were fixed and minor feedback about the wording of the questions was addressed, we proceeded to widely deploy the survey.

## *Results*

**Data Exclusion, Cleansing and Reweighting**

From a total of 156,614 answers, we eliminated all answers with blank or invalid postal codes. Moreover, we only analyze responses with non-blank answers related to age, gender, province and profession (including those who reported not working), yielding a final dataset of 141,865 answers.

Thus, we report the results of analyzing these 141,865 answers collected between 8:00pm GMT of March 28th and 11:59pm GMT on April 2nd. With such large sample, this survey is one the largest citizens' surveys on COVID-19 and the largest in Spain published to date.

All questions are binary or categorical. Thus, we report the percentage of participants who selected each response. Because our gender, age, geographic location and profession distributions were not proportional to those of the general population of Spain, we computed a weighting factor, such that the resulting sample had similar demographic, geographic and profession distributions as those of Spain as reported by the National Institute of Statistics (INE). To reduce biases, we used the reweighted data for all statistical inferences. The user and home situation statistics presented in the next section correspond to the raw data without reweighting. However, the rest of sections regarding the statistical analysis of questions Q8 to Q24 correspond to analyzing the reweighted data.

**Statistical Analyses**

The sampling error, after reweighting the samples, is ±0.43. This small sampling error, due to the large sample, yields a narrow confidence interval of p ± 0.8428 for a 95% confidence level for all proportions reported.

We use the Z test to compare two proportions, considering that the data comes from a survey and as such, the variance of each proportion is different to that of an infinite population test. We use a Chi-squared test to compare the independence between two questions [27]. Differences between answers greater than 0.85 are statistically significant with $P < 0.001$.

We measure the association between nominal variables using Cramer's V for RxC tables and Pearson's phi for 2x2 tables [28] . We use weighted logistic regression in order to compute the Odds Ratio for a multivariate model using a quasi-Binomial distribution family [29].

**User Statistics and Home Situation (questions Q1-Q7)**

Figure 2 displays the demographic information of the respondents: 59.8% were female. In terms of age, we received between 3,324 (age<20) and 38,726 (age between 41-50) answers for each age group.

**Figure 2.** Demographic (age and gender) distribution of the participants

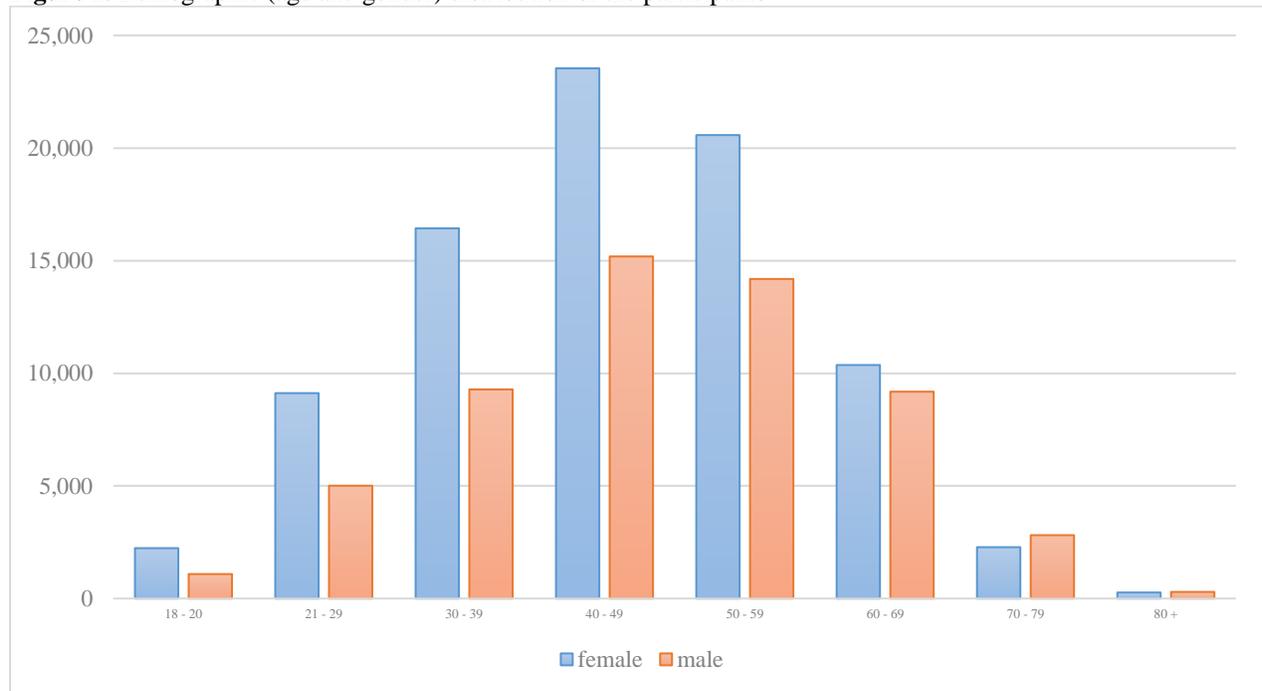

Geographically, most respondents were from the Valencian region (71.9%, N=141,865). However, there were also many answers from other regions of Spain including 10,365 answers from Madrid and 5,691 from Catalonia, as shown in Figure 3.

Given the gender, age and location biases in the raw data, we reweighted the data to match the distribution of the Spanish population according to the latest census [30].

Almost all respondents (98.8%, N=141,807) lived in an apartment (65.6%) or a single-family home (33.1%). Most of the participants lived in a home with 2 (30%), 3 (26.0%) or 4 (27.0%) people, which is consistent with Spain's

demography (N=141,865).

The rest of the reported statistics correspond to analyzing the reweighted sample to match in gender, age, province, and profession the distribution in Spain according to the latest data published by the Spanish National Institute of Statistics (INE).

Given that COVID-19's fatality rates are largest for the elderly [31], we analyze the age distribution of the homes with older adults: 11.8% of respondents under the age of 50 lived with an older adult (age>60) and 19.9% of respondents lived in homes inhabited only by the elderly. Inter-generational homes are particularly important for the transmission of COVID-19 [32].

**Figure 3.** Heat map of survey answer location (generated via ArcGIS)

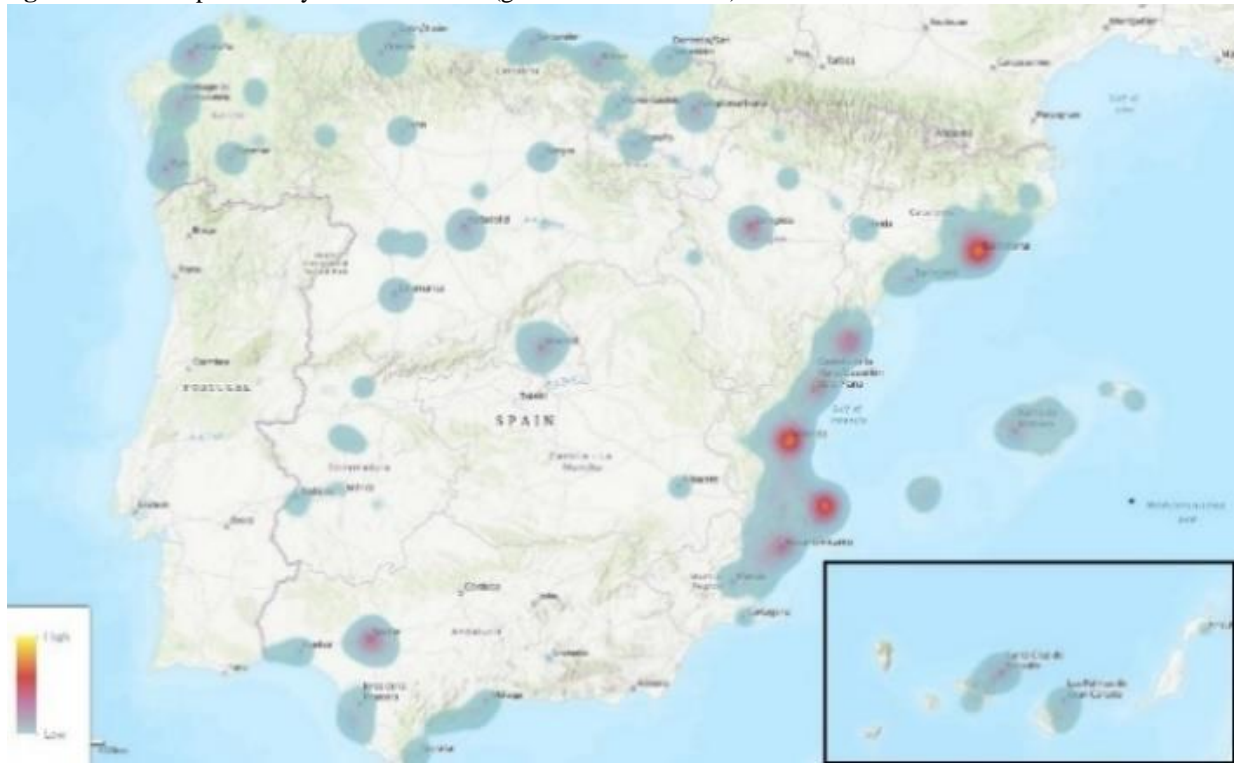

**Social Contact Behavior (questions Q8-Q14)**

With respect to social contact behavior with confirmed COVID-19 patients (Q8), 17.3% of respondents reported having had close contact with a person who was infected with coronavirus (N=140,008). The most common social context was a co-worker (6.2%), a household member (6.1%) or a friend/relative (5.4%). Interestingly, a gender-centric analysis of the answers to this question revealed a significant ($P<0.001$) difference between male / female respondents that had been in close contact with a patient: 60.7% of the respondents were female vs 39.3% male. This large difference is partially due to the larger percentage of women (72.5%) who work in the healthcare sector vs men (27.5%) in Spain [30].

When asked if an outside person regularly visited the home (Q10), we identified a significant difference ($P<0.001$) between older adults (age>70) and younger respondents (N=141,365): 21.2% of older respondents regularly had a person coming to their home vs only 13.6% in the case of younger adults (age< 60). This is an important finding as special measures might need to be taken to protect the 21.2% of older adults who regularly receive external people in their homes.

Respondents left their homes during the social distancing period for a variety of purposes (Q11) as shown in (Figure 4) (N=140,686): covering basic needs (supermarkets and pharmacy) was the most common reason, reported by 47.8% of respondents, followed by going to work (31.3% of respondents). We identified statistically significant differences ($P<0.001$) regarding age and gender. Older respondents (age>60) were more likely than younger participants (age< 60) to stay entirely at home (14.9% older vs 7.6% for younger), and to leave their home to go to the pharmacy (11.5% vs 10.8%) and newspaper stand (9.7% vs 3.9%).

Conversely, younger respondents (age<60) were more likely to leave their home to help others than older respondents (age>60) (81% vs 71.8%). Interestingly, the youngest respondents (18-29 years, N=17,416) were also much more likely to stay entirely at home vs respondents over 30 (23.1 % vs 8.2%).

Regarding gender, among all female respondents 14.8% reported not leaving the home vs 6.5% among male respondents. This difference was statistically significant ($P<0.001$). The opposite significant pattern is found with respect to leaving the home to go to work, where 26% of all female participants vs 36.7% of all male respondents selected this option.

The main means of transportation (Q12) used by respondents was individual, 84.5% (by foot, individual car, motorcycle, scooter) vs shared, 5.9% (public transport, shared car, taxi). In this question, we observe the same gender patterns as in Q11: among female respondents, 13% reported not leaving the home vs 6.2% among male respondents. Moreover, shared transportation means were more likely among female respondents (6.6%) vs their male counterparts (5.1%). These differences are statistically significant ($P<0.001$, N=140,308).

**Figure 4.** Reasons for leaving the home by gender and age

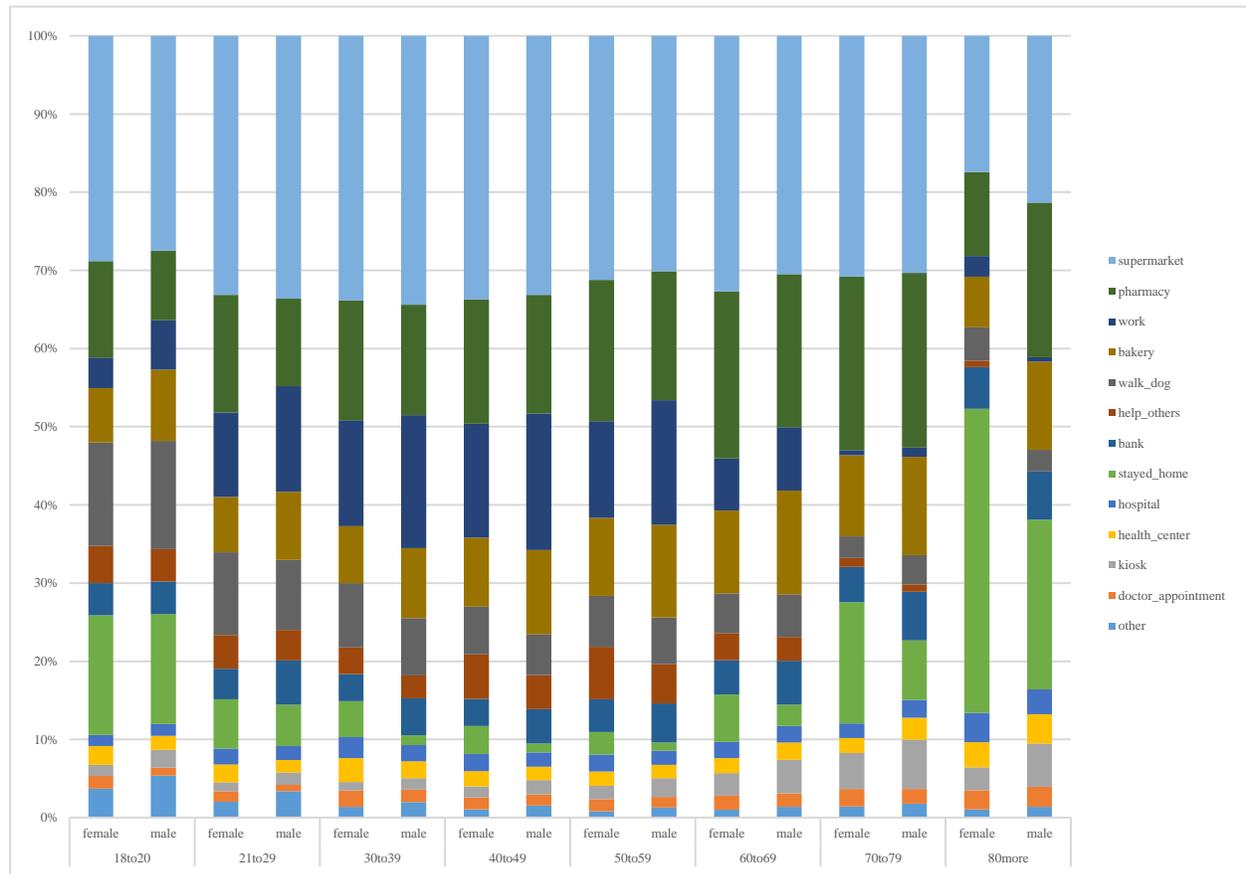

The last two questions in this section (Q13 and Q14) concern the personal experience of respondents regarding the containment measures. Most respondents (46.6%, N=141,481) believed that the government should implement more measures to contain the pandemic and only 2.1% thought that the measures are too severe. There was a significant difference ($P<0.001$) in the support of the measures by age group. Despite being at a lower risk of death, 50% of younger people (age< 60) believed measures should be stronger, vs 37.1% older people (age>60).

Again, there was a significant gender difference in the opinions of respondents. Among female respondents, 48.1%

believe that the government should do more vs 45.1% among male respondents. Regarding whether the measures were too much, the pattern is reversed: among female respondents, 1.5% thought that this was the case, vs 2.7% among male respondents. All differences were statistically significant ($P<0.001$, N=141,481).

Question Q14 explores how sustainable citizens consider the social distancing measures to be. Most respondents (44.1%, N=138,155) answered that they could continue in this confined state for one additional month, yet a non-negligible 32.4% reported being able to continue for 3 to 6 months. An interesting gender difference is found for those who responded that they could stay in confinement for 6 months: among female participants, 8% reported this to be the case vs 12.9% among male participants ($P<0.001$, N=138,155). This might be due to the fact that women see their workload increased during the weeks of social distancing and mobility restriction.

**Personal Economic Impact and Workplace Situation (questions Q15-Q19)**

An inevitable consequence of the COVID-19 pandemic is its economic and labor impact. Spain is a country with mostly small businesses, many of which are family owned. Questions Q15 through Q19 aim to shed light on the individual experiences and fears of people regarding their financial and employment situation.

When asked about the economic impact that the COVID-19 crisis is having on respondents' lives (Q15), 43% (N=139,008) felt that the crisis had not yet significantly affected them economically. Moreover, 29.1% report that their employer or company is undergoing financial problems and 7.7% report having lost a significant part of their savings or their job. Among the respondents who had worked in the last month, there were significant differences in the distribution of work activities, as show in (Table 1). The most affected professions include hospitality and construction. The least affected are education and public administration.

Small businesses have so far borne the brunt of the economic impact. For respondents working in larger companies (100+ employees), 80.1% reported that they had not yet been significantly affected (N=24,386), vs only 42.7% of workers at the smallest companies (1-9 workers) being unaffected (N=39,052). Among those working in small companies, 19.4% reported their companies were facing bankruptcy.

**Table 1.** Distribution of jobs between respondents who had or were in danger of losing their job/business vs those who were not (Cramer's V=0.252).

|  | Lost job or business | Not lost job or business |
|---|---|---|
| **Administrative services** | 6.0% | 7.8% |
| **Retail** | 9.3% | 5.3% |
| **Communications** | 1.2% | 1.8% |
| **Construction** | 18.7% | 7.3% |
| **Domestic service** | 0.2% | 1.5% |
| **Education** | 3.1% | 14.2% |
| **Entertainment/Arts** | 0.8% | 0.7% |
| **Essential services** | 2.2% | 9.7% |
| **Finance** | 1.3% | 4.8% |
| **Food production** | 3.2% | 3.8% |
| **Health and social services** | 1.8% | 2.8% |
| **Hospitality** | 29.3% | 13.8% |
| **Manufacturing** | 6.2% | 4.2% |
| **Other** | 8.8% | 7.4% |
| **Professional/technical/science** | 1.5% | 2.2% |
| **Public administration** | 0.3% | 5.3% |
| **Sanitation** | 2.7% | 4.4% |
| **Transport** | 3.2% | 2.8% |

Again, there is a gender-based statistically significant difference (*P*<0.001, N=139,008): among female respondents, 3.3% reported not being able pay their mortgage and/or pay for food vs 2.4% among male respondents. In terms of having lost their jobs and/or savings, this option was selected by 8.3% among female participants vs 5.9% among male respondents. These figures paint a worrisome picture of the economic impact of the pandemic.

With respect to the labor situation of our respondents (Q16), the majority (71.2%, N=141,865) reported working in the last month. A small fraction, 5.9% of respondents were students.

Question Q17 focused on whether respondents had gone to work in the last week. The answers are split between the three available options: 38.3% did not go to work, 28.7% tele-worked and 33% went to work (N=98,740).

Statistically significant gender differences (*P*<0.001, N=98,740) are observed regarding working participants who: (1) did not go to work (42% among female participants vs 34.9% among male participants) and those who (2) did go to work (29.1% among female participants vs 36.6% among male participants). No significant gender difference was found for those who tele-worked (28.9% among female participants vs 28.5% among male participants). In sum, female workers were significantly more likely to stay home than male workers.

Moreover, we find that economic impact is a key factor in determining resilience towards the confinement measures. To explore the relationship between resilience, economic impact and age, we built a multivariate weighted logistic regression model with *resilience* as dependent variable (answers from Q14, divided into two values: resilience = 0, corresponding to answering that "at most I could continue in confinement for one week" and resilience = 1, corresponding to answering that "I could continue in confinement for longer than one week"). As a covariate variables, we used sex, age and the answers to question Q15 (economic impact). The logistic regression model reveals a clear impact of severe economic damage on resilience: those who report not having enough money to buy food have on average more than twice the probability of reporting zero resilience (i.e. "at most I could continue in confinement for one week"), (OR=2.23, 95%CI [1.81, 2.77]) and those who report being unable to pay their mortgage are on average 1.54 times more likely to also report zero resilience (OR=1.54; 95%CI [1.29, 1.83]). Age also matters: according to the model, respondents aged < 21 have on average over twice the probability to report zero resilience than those aged 21 years and older (OR=2.06, 95%IC [1.73, 2.45]).

**Health State (questions Q20-Q24)**

Finally, questions Q20-Q24 asked respondents about their health. Regarding risk factors, we obtained a similar split between those who belonged to one of the risk groups (48.3%) vs not (46.9%), N=135,583. In addition, 4.9% of respondents were healthcare workers.

Question Q21 aimed to evaluate the ability of respondents to isolate themselves from family members were they to be diagnosed with coronavirus. This is an important question given the relevance of implementing effective quarantine measures during the control phase of the pandemic (after the peak of infections is reached). Whereas 72.3% of respondents reported having the ability to properly isolate themselves, a non-negligible 27.7% of respondents acknowledged not having the necessary resources to implement a proper quarantine in place (N=141,313).

A gender-based analysis reveals statistically significant (*P*<0.001) differences between genders: among female participants, 29.1% reported not having the appropriate infrastructure to isolate themselves when compared to 26.2% among male participants.

In terms of age, 19.8% of respondents aged 80 and older reported not being able to properly isolate themselves in the case that a quarantine was needed, probably because they need assistance in their activities of daily living. It is also notable that all respondents in age groups below 50 years old report not having the appropriate quarantine resources in over 34.9% of cases. This might be due to the presence of children in the home. Indeed, 41.1% of adults with children (N=28,139) vs 28% of adults without children (N=67,659, *P*<0.001) report not being able to properly isolate themselves. Among those living with the elderly (N=15,124), 10.8% reported not having appropriate quarantine infrastructure at home.

To shed light on the percentage of the population that might be infected with coronavirus, Q22 asked respondents if they currently had any of the COVID-19 related symptoms that were unusual for them: 16.8% of respondents reported having at least one of the relevant COVID-19 symptoms (N=136,386), and 7.1% reported having at least one of the more severe symptoms (fever, cough, and difficulty breathing). With regard to gender, a larger percentage of women (19%) vs men (14.5%) reported having symptoms. This difference is statistically significant ($P<0.001$). The age group who most reported having symptoms was the 30-39 year age group (20.9%, N=24,839).

Finally, when asked for whether respondents had been tested for coronavirus (N=138,023), 87.4% felt they didn't need to be tested; 6.1% were told by their doctor they should be tested, but were told no tests were available; 0.7% had tested negative; 0.3% had tested positive, and 0.2% were waiting for their results, resulting in an overall test rate of 1.2%, as shown in (Table 2). We found statistically significant ($P<0.001$) differences between those who exhibited COVID-19 symptoms (difficulty breathing, dry cough and fever) and those who didn't and their answers regarding testing: 93.1% of those who didn't have symptoms considered testing not necessary vs only 58.1% for those who had symptoms.

**Table 2.** Testing needs, depending on the presence of symptoms (all differences between the symptoms/no symptoms groups are statistically significant, $P< 0.001$, and Cramer's V=0.327)

|  | Severe symptoms | Non-severe/no symptoms |
|---|---|---|
| **Negative** | 2.3% | 0.6% |
| **No need** | 58.1% | 93.1% |
| **No test available** | 32.5% | 4.9% |
| **Positive** | 3.9% | 0.2% |
| **Waiting for results** | 1.2% | 0.1% |
| **No want for caretaker** | 2.1% | 1.1% |

When looking at Q8 (whether respondents had had close contact with an infected individual) together with Q23 (whether they had been tested for COVID-19 and the results of the test), we identify an interesting pattern. Among those who had tested positive and answered Q8 (N=414), 80.9% had had close contact with a known infected individual; of these, 32.4% had been through a friend or relative; 26.6% through a patient (they were healthcare workers); 11.1% at work and only 1.7% through a client. This means that in a remarkably high percentage of cases, respondents with COVID-19 knew their likely source of infection. This finding is partly explained by the fact that the survey was answered during a period of reduced mobility.

Finally, we carried out a multivariate weighted logistic regression analysis to study the relationship between testing positive and being able to self-isolate (Q21), gender and sex. We found a triple interaction between these variables. Females aged 70 or older who report not being able to properly isolate have on average almost twice the probability of testing positive than otherwise (OR= 1.91, 95%CI [1.18, 3.073]).

**Prevalence**

One of the principal goals of the survey was to be able to rapidly estimate the prevalence of COVID-19 in the Spanish population. Given the lack of tests during the initial phases of the pandemic and the large percentage of mildly symptomatic or asymptomatic patients [33], the survey could be a useful tool to quickly assess the percentage of infected individuals in the population. Thus, we created a linear regression model to infer prevalence of COVID-19, using as independent variables a subset of the reported symptoms (Q22), whether the household already had an infected member (Q8), gender and whether the participant's age was older than 70 years old. Our target variable was given by Q24: we selected those answers corresponding to participants who reported having tested positive (coded as 1) /negative (coded as 0) for COVID-19 (N=1,345).

The obtained value from the regression function is converted into a probability using the logistic function $(\exp(x)/(1+\exp(x)))$. Moreover, we performed feature selection among all possible symptoms to select those that yielded the best performing model. The variables and parameters of such a model are given in (Table 3). The final model has a sensitivity of 0.77 and a specificity of 0.80. Although we experimented with more sophisticated machine learning models, for the purpose of this paper we wanted show that we could arrive at a reasonable estimate using a simple and easily reproducible method.

Based on the model with the coefficients below, we estimated the number of infected individuals among all the respondents to the survey. Geographically, we aggregated the results by the 17 autonomous regions of Spain, making it easier to compare with official data, since each autonomous region has its own health care system, and official figures are always reported by autonomous region.

Table 3. Linear regression coefficients on responses that included only a positive or negative test result.

| Variable | Estimate | Std Error | T value | Pr(>|t|) |
|---|---|---|---|---|
| (Intercept) | 0.12510 | 0.01636 | 7.647 | 4.05e-14 |
| Member of home infected | 0.28555 | 0.03137 | 9.103 | < 2e-16 |
| Fever | 0.18569 | 0.03724 | 4.986 | 7.01e-07 |
| Dry cough | 0.05834 | 0.02808 | 2.078 | 0.037937 |
| Productive cough | -0.09785 | 0.03676 | -2.662 | 0.007875 |
| Muscle Pain | 0.07208 | 0.03603 | 2.000 | 0.045670 |
| Loss of sense of smell | 0.45410 | 0.03409 | 13.319 | < 2e-16 |
| Age > 70 | 0.17487 | 0.06785 | 2.577 | 0.010068 |
| Male | 0.08038 | 0.02306 | 3.485 | 0.000509 |

On May 13th, 2020, more than a month after we had carried out our analysis, the Spanish National Statistics Institute (INE) published the initial results of a nationwide seroprevalence study performed between April 27th and May 11th [22]. Using the data published from this study, we were able to empirically validate our prevalence estimations, depicted in (Table 4). Regression estimates that are within the confidence intervals of the estimated prevalence by the seroprevalence study are highlighted in bold.

Since the survey only asked about current symptoms, the seroprevalence survey would have covered additional cases detected after the end of the survey on April 3rd and cases that had recovered by March 28th. However, we did not find that there was a significant underestimation of cases, which could be due to most cases occurring near the time period covered by our survey. Using official statistics to estimate the distribution of the number of cases was not possible, since by May 11th Spain had only recorded 268,729 confirmed cases, that is only 11% of the 2,350,000 positive cases estimated by the seroprevalence study.

Table 4. Comparison of seroprevalence study and regression estimate using the survey.

| Autonomous Community | N | Regression estimate [CI = 95%] | Seroprevalence Survey [CI = 95%] |
|---|---|---|---|
| Andalucía | 5,691 | 4.4% [±0.5] | 2.7% [2.2-3.2] |
| Aragón | 1,463 | **3.1% [±0.9]** | 4.9% [3.8-6.3] |
| Asturias | 655 | **4.0% [±1.5]** | 1.8% [1.3-2.5] |
| Balearic Islands | 1,222 | **4.2% [±1.1]** | 2.4% [1.6-3.5] |
| Canarias | 1,052 | **3.0% [±1.0]** | 1.8% [1.1-2.8] |
| Cantabria | 497 | **4.6% [±1.8]** | 3.2% [2.1-5.0] |
| Castilla y León | 1,994 | **6.1% [±1.0]** | 7.2% [6.3-8.1] |
| Castilla-La Mancha | 3,469 | **10.4% [±1.0]** | 10.8% [9.3-12.4] |
| Catalonia | 5,088 | **4.8% [±0.6]** | 5.9% [4.9-6.9] |
| Valencia | 102,021 | 3.4% [±0.1] | 2.5% [1.9-3.2] |
| Extremadura | 656 | **4.4% [±1.6]** | 3.0% [2.2-4.1] |
| Galicia | 2,257 | **2.6% [±0.7]** | 2.1% [1.7-2.6] |
| Madrid | 10,365 | 8.8% [±0.5] | 11.3% [9.8-13.0] |
| Murcia | 3,566 | 3.2% [±0.6] | 1.4% [0.8-2.4] |
| Navarra | 580 | **5.5% [±1.9]** | 5.8% [4.3-7.6] |
| País Vasco | 1,007 | **3.9% [±1.2]** | 4.0% [3.1-5.2] |
| Rioja, La | 220 | **5.0% [±2.9]** | 3.3% [2.4-4.4] |

**Principal Findings**

Through the survey answers we identify several patterns and implications for the design of public policies in the context of the COVID-19 pandemic.

First, our work highlights the value of involving citizens and carrying out large-scale online surveys for a quick assessment of the population's situation and perceptions during a pandemic. We were overwhelmed by the extremely positive response of citizens to the survey. Mayors in large and small towns got involved and shared it with their employees and citizens; professional and civic associations disseminated it among their members; individuals advertised it among their contacts, and a few media organizations gave it visibility via articles and posts. Most respondents were enthusiastic and supportive of the initiative, yet also asked for the results to be shared as soon as possible. This outstanding response by people might reflect a societal need to have more information about the impact of COVID-19 in our lives, but is also a wonderful example of citizen's science and people's willingness to help by contributing with their answers to achieve a more data-driven decision-making processes. While the sample has some biases, we were able to compensate for them via reweighting.

Second, we empirically corroborate the impact that close contacts play in the transmission of the disease. Over 16% of respondents reported having had close contact with someone who was infected by the coronavirus. This percentage was much higher (80.9%) among those who had tested positive for COVID-19. This increases the likelihood that those testing positive were infected by someone they knew and had close contact with, rather than, for example, a random infected stranger in a supermarket. This finding could have implications for contact tracing strategies.

Third, gender matters. Several statistically significant differences were found between male and female respondents, with a clear pattern of placing women in situations of higher vulnerability or exposure when compared to men. As in other aspects of society, gender-based differences exist in the context of a pandemic. It is a socially important factor that needs to be considered.

Fourth, age also matters. We identified statistically significant differences in the social contact behavior questions between older participants (age > 70) and younger participants (age < 60). Older respondents were more likely to stay at home, and to leave their house to go to the pharmacy and newspaper stand. There were also different aged-based attitudes towards the containment measures: younger participants were significantly more supportive of stronger measures than older participants while they were more likely to report not being able to stand the confinement any more (4.9%) vs older adults (0.77%). We also found associations between age, gender, the ability to self-isolate and the probability of testing positive: older females without the capacity of isolate themselves were almost twice as likely to test positive than otherwise.

Citizens demanded more measures, as over 46% of respondents were supportive of implementing additional social distancing measures. This result might reflect the worry in people's minds regarding the exponential progression of the pandemic and the lack of clear signs of flattening the curve at the time of answering the survey.

Moreover, citizens were willing to sustain social distancing for a month or more. The citizens' solidarity with the measures is reflected by the fact that over 32% of respondents reported being able to stay at home for three to six additional months.

The economic impact of the pandemic is evident, particularly for those working in small companies, 19.4% of which were reported to be facing bankruptcy. Moreover, over 47.3% of participants who worked in small companies reporting having been impacted by the pandemic. In terms of professions, hospitality, construction and retail were the most affected. Hospitality represents 6.2% of the Spanish GDP [34] and construction 5.6% [35].

Among those who were working, roughly 29% of respondents reported tele-working and one third leaving the home to go to work. The tele-work figure is lower than in other countries. For example, in the US, it is estimated that 56-62% of the workforce could work remotely. Moreover, on March 31st, the Government established labor mobility restrictions for all non-essential professions. Given that 71.2% (N=141,865) of respondents reported having worked in the last month, our expectation is that this ~23% of the population will be impacted by such measures. Regarding workplace infections, we found that 11.1% of those who tested positive (and did not work in the healthcare sector) had had close contact someone at work who had tested positive for COVID-19.

Quarantine infrastructure might be needed, as over 27.7% of respondents reported not having the appropriate infrastructure to isolate themselves at home. Effective quarantine measures for asymptomatic or lightly symptomatic patients are key to control the spread of the pandemic. Thus, developing the needed infrastructure might be key to slow down the transmission of the disease.

The number COVID-19 infected individuals is certainly larger than what has been officially reported. In our survey, over 16.8% of respondents reported having at least one of the COVID-19 related symptoms and over 7.1% reported having at least one of more severe symptoms (fever, cough and/or difficulty breathing). We show how the prevalence of a rapidly spreading disease, such as COVID-19, can be estimated quite accurately using a citizen survey. From the answers to two of the questions plus demographic information, we built a regression model and report COVID-19 prevalence estimations that are at par with those carried out by a recent seroprevalence study in Spain. Thus, when public policy decisions need to be made rapidly, a citizen-science survey can be deployed rapidly and collect results within hours.

Finally, in the context of Spain, our survey revealed a lack of tests. In terms of testing capabilities, over 6.1% of respondents reported not having been able to do the test despite their doctor's recommendation. Moreover, a significant difference was found between those who had the more severe COVID-19 symptoms (32.5%) and those who did not regarding their attitudes towards the need for testing. Given such a large percentage of the population with symptoms, it is evident that there was a need for many more tests.

**Limitations**

While the sample size in our study is large, our methodology is not exempt of limitations. First, there is a potential selection bias, given that all participants volunteered to fill out the survey without any incentive. In our analysis, we have corrected for gender, age, location, and profession biases via reweighting, but this fact should be noted. Next, this is an in-the-wild study and thus people could provide untruthful answers. We addressed this limitation by filtering entries without proper zip codes and entries that had inconsistencies in them.

## Conclusions

The COVID-19 pandemic is undoubtedly impacting the lives of citizens worldwide. While there is abundant data regarding the number of reported cases, hospitalizations and intensive care patients and deaths, there is a scarcity of data about the individual experiences of people, their personal, financial and labor situations, their health state and their resilience towards the confinement measures. This paper reports the first results of analyzing a large-scale, rich dataset of self-reported information regarding the social contact, economic impact, working situation and health status of over 140,000 individuals in Spain. It is probably one of the largest population surveys in a single country carried out in the context of an infectious disease pandemic.

The data is extremely rich and multi-faceted. Thus, it offers numerous avenues of future work and deeper analysis according to different dimensions, including location (at a zip code level) which we have not covered in this paper.

We have launched successive versions of the *Covid19Impact* survey [26] in consecutive weeks throughout the COVID-19 pandemic, to assess the pulse of the virus from the perspective of citizens over time and assess changes in people's situations and perceptions regarding the pandemic.

**Acknowledgments**

We thank the thousands of citizens who volunteered to fill out this survey and shared it with their contacts. Their generosity and enthusiasm have enabled this valuable dataset to be collected. Kristof Roomp would like to thank his employer, Microsoft, for letting him volunteer his time to help on this effort. This project has been carried out in collaboration with the Valencian Government of Spain.

**Authors' Contributions**

NO and KR conceptualized the study, interpreted and analyzed the data, drafted the manuscript, and provided supervision. KR deployed the survey and parsed the responses. XB reweighted and analyzed the data. Kirsten Roomp conceptualized the study, carried out the literature review and provided feedback to the manuscript. All authors approved the final version of the manuscript.

**Conflicts of Interest**

None declared.

**Appendices 1 and 2**

The online survey questionnaire and answers.

**Abbreviations**

API: application program interface
CDC: Centers for Disease Control and Prevention
COVID-19: coronavirus disease
WHO: World Health Organization

## Appendix 1 – Survey Questions *(Translated from Spanish)* http://covid19impactsurvey.org

| | | |
|---|---|---|
| Start | Consent | I am an adult and I consent to taking this survey |
| | | I am not an adult, or I do not consent taking this survey *(skip to end)* |
| | Q1 In which country are you presently in? | Spain and other Latin American countries |
| Basic data | Q2 What is your age range? | 18-20 |
| | | 21-29 |
| | | 30-39 |
| | | 40-49 |
| | | 50-59 |
| | | 60-69 |
| | | 70-79 |
| | | 80 or more |
| | Q3 What is your gender? | Male |
| | | Female |
| | Q4 Postal code | Entered as text |
| Home situation | Q5 Type of home | Single Family |
| | | Apartment |
| | | Old age home |
| | | Home for disabled people |
| | | Prison/Jail |
| | | Hotel |
| | | Other shared accommodation (monastery, etc.) |
| | | Camping |
| | | Homeless |
| | | Other |
| | Q6 Number of people in home (including you) | 1 |
| | | 2 |
| | | 3 |
| | | 4 |
| | | 5 or more |
| | Q7 Age(s) of people in your home *(check all that apply)* | 10 or less |
| | | 11-20 |
| | | 21-29 |
| | | 30-39 |
| | | 40-49 |
| | | 50-59 |
| | | 60-69 |
| | | 70-79 |
| | | 80 or more |
| Social contact in the last two weeks | Q8 Have you had physical contact with someone diagnosed with coronavirus? *(check all that apply)* | None that I know of |
| | | Member of household |
| | | Family outside household |
| | | Friend |
| | | Coworker |
| | | Cleaning staff/nurse/etc. |
| | | Patient (in case of medical staff) |
| | | Client/Customer |
| | Q9 If you have children, are they taken care of by someone outside the home (grandparents, neighbors, etc.)? | Yes |
| | | No |
| | | I don't have children |
| | Q10 Does anyone who doesn't live in your home regularly enter (cleaner, nurse, caretaker, etc.)? | Yes<br>No |
| | Q11 For what activities do you leave your home?<br>*(check all that apply)* | Go to hospital<br>Go to a doctor's appointment<br>Go to a health care center (blood test, anticoagulants, etc.)<br>Go to work<br>Go to supermarket<br>help someone that lives outside your home<br>Go to the bank<br>Go to the pharmacy<br>Go to the bakery<br>Go to the newspaper stand<br>Walk the dog<br>Other<br>Stayed home the whole time |
| | Q12 What means of transport do you use?<br>*(check all that apply)* | Walk<br>Motorcycle<br>Car (individual)<br>Car (shared)<br>Bike/scooter<br>Public transport (bus, train. etc.)<br>Taxi/Uber/etc.<br>Stayed home |
| | Q13 Do you believe that the measures the government have taken are enough to contain the spread of coronavirus? | No, should be stricter<br>Yes, are about right<br>Yes, but are too strict<br>Prefer not to respond<br>Don't know |
| | Q14 If you are currently confined to not leaving your home, how much longer can you stand it? | 0 days, I can't stand it anymore<br>1 week<br>2 weeks<br>1 month<br>2 months<br>6 months |
| Economic impact | Q15 What kind of economic impact has the coronavirus had on you?<br>(check all that apply) | No or little impact<br>I lost my job<br>I lost my savings<br>I can't pay my mortgage anymore<br>I can't afford to buy food<br>My business is in danger of bankruptcy |
| | Q16 Have you gone to work in the last month? | Yes<br>No<br>No, I'm a student |
| Workplace *(skip unless the previous answer was yes)* | Q17 Have you gone to work in the last week? | Yes<br>No<br>No, but I'm teleworking |
| | Q18 How many people work at your place of work? | 1-9<br>10-99<br>100+ |
| | Q19 What is your main type of work? | Essential services (police, fireman, doctor)<br>Retail large/small<br>Manufacturing<br>Health and social services |

| | | Hospitality |
| --- | --- | --- |
| | | Education |
| | | Government or defense |
| | | Construction |
| | | Transport |
| | | Administrative assistant and similar |
| | | Professional, technical, scientist |
| | | Farming, fishing or other food production |
| | | Press or communication |
| | | Domestic care |
| | | Financial |
| | | Arts, entertainment, recreation |
| | | Sanitation, cleaning, garbage collection |
| | | Other services |
| Health | Q20 Are you a member of any of these risk groups? *(check all that apply)* | Hypertension Diabetes Cardiovascular disease Respiratory illness Immuno-suppressant Cancer Smoker (current) Smoker (ex) Pregnant Health care worker Not in a risk group I prefer not to answer |
| | Q21 If you were diagnosed with coronavirus, would you be able to isolate yourself from other members in your home? | Yes No |
| | Q22 Do you have any of the following symptoms (more than normal) *(check all that apply)* | Fever Dry cough Productive cough Difficulty breathing Sore throat Headache Muscle pain Loss of sense of smell None of these symptoms I prefer not to answer |
| | Q23 How long have you had these symptoms? | I don't have these symptoms 1 - 3 days 4 - 7 days 8 - 13 days 14 or more days I prefer not to answer |
| | Q24 Have you taken the test for coronavirus? | No, but I don't think I need it No, my doctor recommended it but there weren't any tests available Yes, I'm waiting for my result Yes, the result is I have COVID-19 Yes, the result is I don't have COVID-19 I prefer not to answer |

## Appendix 2 – Survey answers -- Univariate tables

**Q2 What is your age range?**

| Q2 | N | % | % weighted | INE[+] % |
|---|---|---|---|---|
| <30 | 17,452 | 12.3 | 13.6 | 13.9 |
| 30-39 | 25,719 | 18.1 | 18 | 18.1 |
| 40-49 | 38,726 | 27.3 | 22.5 | 22.3 |
| 50-59 | 34,762 | 24.5 | 19.8 | 19.7 |
| 60-69 | 19,551 | 13.8 | 14.9 | 14.8 |
| 70-79 | 5,093 | 3.6 | 10.8 | 10.8 |
| 80+ | 562 | 0.4 | 0.3 | 0.3 |
| Total | 141,865 | | | |

INE: Spanish National Institute of Statistics

**Q3 What is your gender?**

| Q4 | N | % | % weighted | INE % |
|---|---|---|---|---|
| Female | 84,819 | 59.8 | 50.8 | 50.9% |
| Male | 57,046 | 40.2 | 49.2 | 49.1% |
| Total | 141,865 | | | |

**Q5 Type of home**

| Q5 | N | % | % weighted |
|---|---|---|---|
| Appartment | 93,060 | 65.62 | 66.28 |
| Camping | 45 | 0.03 | 0.04 |
| Disabled home | 10 | 0.01 | 0.01 |
| Homeless | 29 | 0.02 | 0.02 |
| Hotel | 31 | 0.02 | 0.04 |
| Nursing home | 36 | 0.03 | 0.03 |
| Other | 1,338 | 0.94 | 0.86 |

| Q5 | N | % | % weighted |
|---|---|---|---|
| Other shared home | 255 | 0.18 | 0.23 |
| Prison | 28 | 0.02 | 0.04 |
| Single family home | 46,975 | 33.13 | 32.46 |
| Total | 141,807 | | |

**Q6 Number of people in your home (including yourself)**

| Q6 | N | % | % weighted |
|---|---|---|---|
| 1 | 13,969 | 9.8 | 11.1 |
| 2 | 42,513 | 30 | 31.6 |
| 3 | 36,879 | 26 | 24.2 |
| 4 | 38,265 | 27 | 25.1 |
| 5+ | 10,239 | 7.2 | 8 |
| Total | 141,865 | | |

**Q7* Ages of people in your home (check all that apply)**

| Q7 | N | % | % weighted |
|---|---|---|---|
| <10 | 32,666 | 23.7 | 20.6 |
| 11-20 | 35,646 | 25.9 | 25.2 |
| 21-29 | 28,742 | 20.9 | 19.5 |
| 30-39 | 30,030 | 21.8 | 21.7 |
| 40-49 | 42,766 | 31.1 | 28.9 |
| 50-59 | 44,396 | 32.2 | 30.3 |
| 60-69 | 26,257 | 19.1 | 19.8 |
| 70-79 | 8,934 | 6.5 | 11.9 |
| 80+ | 5,362 | 3.9 | 3.7 |
| Total | 137,704 | | |

* Multiple answer question

**Q8* Have you had physical contact with someone diagnosed with coronavirus?**

| Q8 | N | % | % weighted |
|---|---|---|---|
| Noone | 119,095 | 85.1 | 82.8 |
| Household member | 7,172 | 5.1 | 6.1 |
| Relative | 3,443 | 2.5 | 2.8 |
| Friend | 2,487 | 1.8 | 2.6 |
| Coworker | 7,577 | 5.4 | 6.2 |
| Cleaning person | 683 | 0.5 | 0.7 |
| Sick patient | 3,443 | 2.5 | 2.6 |
| Professional client | 1,219 | 0.9 | 1 |
| Total | 140,008 | | |

**Q9 If you have children, are they taken care of by someone outside the home (grandparents, neighbors, etc…)?**

| Q9 | N | % | % weighted |
|---|---|---|---|
| No | 82,479 | 59.2 | 56.7 |
| No kids | 48,341 | 34.7 | 37.4 |
| Yes | 8,535 | 6.1 | 5.9 |
| Total | 139,355 | | |

**Q10 Does anyone who does not live in your home regularly enter your house (e.g. cleaner, nurse…)?**

| Q10 | N | % | % weighted |
|---|---|---|---|
| No | 121,657 | 86.1 | 85.3 |
| Yes | 19,708 | 13.9 | 14.7 |
| Total | 141,365 | | |

**Q11\* For what activities do you leave your home? (check all that apply)**

| Q11 | N | % | % weighted |
|---|---|---|---|
| Hospital | 7,265 | 5.1 | 5.3 |
| Doctor appointment | 5,333 | 3.8 | 3.8 |
| Health center | 7,202 | 5.1 | 4.7 |
| Work | 44,593 | 31.4 | 32.0 |
| Supermarket | 112,567 | 79.3 | 79.1 |
| Help others | 15,836 | 11.2 | 9.5 |
| Bank | 15,241 | 10.7 | 11.0 |
| Pharmacy | 58,074 | 40.9 | 40.9 |
| Bakery | 41,142 | 24.1 | 23.1 |
| Kiosk | 23,058 | 21.2 | 21.2 |
| Walk dog | 20,440 | 14.4 | 12.8 |
| Other | 12,260 | 8.6 | 8.6 |
| Stayed home | 1,179 | 0.8 | 1.0 |
| Total | 140,686 | | |

**Q12\* What means of transport do you use? (check all that apply)**

| Q12 | N | % | % weighted |
|---|---|---|---|
| Walk | 78,998 | 56.1 | 55.6 |
| Bike | 1,144 | 0.8 | 0.8 |
| Public transport | 3,005 | 2.1 | 2.8 |
| Motorcycle | 2,110 | 1.5 | 2 |
| Car shared | 3,174 | 2.3 | 2.7 |
| Car individual | 77,751 | 55.2 | 53.1 |
| Stayed home | 12,511 | 8.9 | 9.6 |
| Taxi | 852 | 0.6 | 0.7 |
| Total | 140,799 | | |

**Q13 Do you believe that that measures the government have taken are enough to contain the spread of the coronavirus?**

| Q13 | N | % | % weighted |
|---|---|---|---|
| Do more | 65,453 | 49.4 | 50.4 |
| Enough | 36,624 | 27.7 | 27.3 |
| Too much | 2,422 | 1.8 | 2.2 |
| Don't know | 27,899 | 21.1 | 20.1 |
| Total | 141,481 | | |

**Q14 If you are currently confined to not leaving your home, how much longer can you stand it?**

| Q14 | N | % | % weighted |
|---|---|---|---|
| I can't anymore | 1,877 | 1.4 | 1.5 |
| 1 week | 4,108 | 3 | 3.2 |
| 2 weeks | 26,473 | 19.2 | 18.9 |
| 1 month | 61,412 | 44.5 | 44.1 |
| 3 months | 30,134 | 21.8 | 21.9 |
| 6 months | 14,151 | 10.2 | 10.5 |
| Total | 138,155 | | |

**Q15* What kind of economic impact has the coronavirus had on you? (check all that apply)**

| Q15 | N | % | % weighted |
|---|---|---|---|
| None | 93,132 | 65.6 | 63 |
| Lost job | 9,322 | 6.6 | 8.1 |
| Lost savings | 10,634 | 7.5 | 7.8 |
| Can't pay mortgage | 10,449 | 7.4 | 7.9 |
| No food | 3,644 | 2.6 | 2.6 |
| Company bankrupt | 11,039 | 7.8 | 9.2 |
| Employer bankrupt | 956 | 0.7 | 0.9 |
| Total | 141,865 | | |

**Q16 Have you gone to work in the last month?**

| Q16 | N | % | % weighted |
|---|---|---|---|
| No | 36,804 | 25.9 | 22.8 |
| No, I'm student | 6,640 | 4.7 | 5.9 |
| Yes | 98,421 | 69.4 | 71.2 |
| Total | 141,865 | | |

**Q17 Have you gone to work in the last week?**

| Q17 | N | % | % weighted |
|---|---|---|---|
| No | 32,150 | 32.6 | 38.3 |
| Teleworking | 32,787 | 33.2 | 28.7 |
| Yes | 33,803 | 34.2 | 33 |
| Total | 98,740 | | |

**Q18 How many people work at your place of work?**

| Q18 | N | % | % weighted |
|---|---|---|---|
| 100+ | 24,386 | 25 | 25.1 |
| 10-99 | 33,947 | 34.9 | 33.4 |
| 1-9 | 39,052 | 40.1 | 41.5 |
| Total | 97,385 | | |

**Q19 What is your main type of work?**

| Q19 | N | % | % weighted |
|---|---|---|---|
| Admin services | 5,327 | 5.4 | 7.6 |
| Retail large/small | 7,164 | 7.3 | 5.7 |
| Press or communication | 3,549 | 3.6 | 1.8 |
| Construction | 2,723 | 2.8 | 8.7 |
| Domestic care | 883 | 0.9 | 1.4 |
| Education | 16,879 | 17.1 | 12.9 |
| Entertainmnet | 2,023 | 2.1 | 0.7 |
| Essential services (police, fireman, doctor) | 7,692 | 7.8 | 8.9 |
| Financial | 3,064 | 3.1 | 4.4 |
| Farming, fishing or other food production | 1,633 | 1.7 | 3.8 |
| Health and social services | 7,425 | 7.5 | 2.7 |
| Hospitality | 3,525 | 3.6 | 15.7 |
| Manufacturing | 4,507 | 4.6 | 4.4 |
| Other services | 12,296 | 12.5 | 7.5 |
| Professional, technical, scientist | 8,475 | 8.6 | 2.1 |
| Government or defense | 8,607 | 8.7 | 4.7 |
| Sanitation, cleaning, garbage collection | 628 | 0.6 | 4.2 |
| Transport | 2,261 | 2.3 | 2.8 |
| Total | 98,661 | | |

**Q20\* Are you a member of any of these risk groups? (check all that apply)**

| Q20 | N | % | % weighted |
|---|---|---|---|
| Hypertension | 17,387 | 12.3 | 14.4 |
| Diabetes | 5,133 | 3.6 | 4.2 |
| Cardiovasular | 4,677 | 3.3 | 4.1 |
| Respiratory | 8,421 | 5.9 | 6.3 |
| Immunocompromised | 2,926 | 2.1 | 2 |

| Q20 | N | % | % weighted |
|---|---|---|---|
| Cancer | 2,674 | 1.9 | 2.1 |
| Smoker | 42,429 | 29.9 | 30 |
| Exsmoker | 17,324 | 12.2 | 12.5 |
| Pregnant | 1,039 | 0.7 | 0.5 |
| Healthcare worker | 8,000 | 5.6 | 4.6 |
| None | 65,074 | 45.9 | 44.4 |
| Prefer not to say | 5,819 | 4.1 | 4.5 |
| Total | 141,865 | | |

**Q21 If you were diagnosed with coronavirus, would you be able to isolate yourself from other members of your home?**

| Q21 | N | % | % weighted |
|---|---|---|---|
| No | 40,083 | 28.4 | 27.7 |
| Yes | 101,230 | 71.6 | 72.3 |
| Total | 141,313 | | |

**Q22* Do you have any of the following symptoms (more than normal)? (check all that apply)**

| Q22 | N | % | % weighted |
|---|---|---|---|
| Fever | 2,007 | 1.4 | 1.7 |
| Dry cough | 6,835 | 4.8 | 5.2 |
| Productive cough | 6,353 | 4.5 | 4.4 |
| Difficulty breathing | 2,125 | 1.5 | 1.6 |
| Sore throat | 7,924 | 5.6 | 5.3 |
| Headache | 5,780 | 4.1 | 4.2 |
| Muscle pain | 3,629 | 2.6 | 2.8 |
| Loss of smell | 2,793 | 2 | 2.6 |
| None | 113,888 | 80.3 | 79.2 |
| Prefer not to say | 5,268 | 3.7 | 4.5 |

| Q22 | N | % | % weighted |
|---|---|---|---|
| Total | 141,865 | | |

**Q23 How long have you had these symptoms for?**

| Q23 | N | % | % weighted |
|---|---|---|---|
| 1-3 | 6,045 | 4.7 | 4.8 |
| 4-7 | 5,794 | 4.5 | 4.8 |
| 8-13 | 4,348 | 3.3 | 3.6 |
| 14+ | 5,085 | 3.9 | 4.1 |
| No symptoms | 107,792 | 83.0 | 82.2 |
| Prefer not to say | 262 | 0.6 | 0.5 |
| Total | 129,064 | | |

**Q24 Have you taken the test for coronavirus?**

| Q24 | N | % | % weighted |
|---|---|---|---|
| Prefer not to say | 5,164 | 3.7 | 3.6 |
| Yes, result is I do not have COVID-19 | 919 | 0.7 | 0.7 |
| No, my doctor recommended but there are no tests available | 8,412 | 6.1 | 6.9 |
| No, but I do not think I need it | 121,323 | 87.9 | 86.9 |
| No, but would want to as I am a caretaker | 1,518 | 1.1 | 1.2 |
| Yes, the result is I have COVID-19 | 426 | 0.3 | 0.5 |
| Yes, I am waiting for my results | 261 | 0.2 | 0.2 |
| Total | 138,023 | | |